\documentclass[usenatbib]{mn2e}
\input{psfig.sty}
\usepackage{graphicx}

\def\beq{\begin{equation}}
\def\enq{\end{equation}}
\def\ms{$M_\odot$}
\title{Variations and Correlations in Cyclotron Resonant Scattering Features of Vela X-1 Studied by INTEGRAL}

\author[Wei Wang]{Wei Wang\thanks{E-mail: wangwei@bao.ac.cn} \\
National Astronomical Observatories, Chinese Academy of Sciences,
Beijing 100012, China}

\begin{document}

\maketitle

\begin{abstract}
Long-term hard X-ray monitoring observations on high mass X-ray binary Vela X-1 from 2003 -- 2011 have been performed by INTEGRAL. I systematically analyzed the average hard X-ray spectra of Vela X-1 from 3 -- 200 keV, with main aims to detect cyclotron resonant scattering features and study their variation patterns with accreting luminosities and orbital phases. The cyclotron scattering lines of Vela X-1 at $\sim 22 - 27$ keV and $49 - 57$ keV are confirmed in the average spectra of Vela X-1. But in the flare states with hard X-ray luminosity higher than $\sim 5\times 10^{36}$ erg s$^{-1}$ (3 -- 100 keV), the fundamental line cannot be detected. This feature suggests that a fan-like beam radiation pattern is expected in high luminosity ranges of Vela X-1. Variations of the cyclotron line energies and depths are discovered, which may change with continuum spectral properties. Both cyclotron line energies show no correlations with X-ray luminosity. The fundamental line energy shows no significant correlations with photon index and cutoff energy. While, the first harmonic energy shows a positive correlation with photon index and exponential cutoff energy. The energy ratio of two cyclotron lines always higher than 2 has the weak correlation with photon index and cutoff energy. These relations support that the X-ray spectral properties of accreting X-ray pulsars are affected by cyclotron resonance scattering. In the case of Vela X-1, the broader and deeper first harmonic would play the main role to cause the spectral variations. The positive correlations between the ratio of line width to energy and the corresponding depth for both two lines support a cylindrical column accretion geometry in Vela X-1.
\end{abstract}


\section{Introduction}

Vela X-1 is a persistently emitting and eclipsing high mass X-ray binary (HMXB) with a B0.5 1b supergiant companion whose distance is determined at $\sim 1.9$ kpc (Nagase 1989). Its X-ray luminosity variability shows an orbital period of 8.96 days (van Kerkwijk et al. 1995). This source contains a neutron star with a pulsation period of $\sim 283$ s (McClintock et al. 1976). The spin period and the derivative changed erratically since its discovery as expected from the wind-fed direct accretion system (Bildsten et al. 1997).

Cyclotron resonance scattering features (CRSFs) can be used to directly determine the magnetic field of the neutron star, and additionally reflect the accretion geometry and emission properties near the surface of the neutron star. An absorption line feature around 55 keV was first reported in Vela X-1 from HEXE observations (Kendziorra et al. 1992). In addition, Makishima \& Mihara (1992) found a possible absorption feature around 32 keV from Ginga data, and Choi et al. (1996) reported the line feature around 23 -- 24 keV also using the Ginga data. Kretschmer et al. (1997) reported two absorption features at $\sim 23$ keV and $\sim 45$ keV using the broadband data from HEXE. With RXTE data, a detailed pulse phase resolved analysis (Kreykenbohm et al. 2002) confirmed the existence of two cyclotron scattering lines at around 25 keV and 55 keV, and also reported its variation with the pulse phase. However, some other people only found the cyclotron scattering feature around 55 keV using the BeppoSAX data (e.g., Orlandini et al. 1998). With recent INTEGRAL data, Schanne et al. (2007) found two possible absorption lines at $\sim 27$ keV and 54 keV. Kreykenbohm et al. (2008) reported the absorption line at $\sim 53$ keV using the INTEGRAL/IBIS data. More recently, detailed analysis of the Suzaku observational data by Maitra \& Paul (2013) and Odaka et al. (2013) reported the detection of CRSFs at $\sim 25$ keV and 50 keV. Thus there exist two cyclotron scattering lines in Vela X-1 with present observations by different missions, but the understanding of variability pattern of CRSFs still requires more observational data and detailed studies.

Previous observations on accreting X-ray pulsars studied the correlations between the CRSFs and luminosities/spectral properties mainly in Be X-ray transients (see Klochkov et al. 2012; Li et al. 2012). But the variation pattern and the possible correlations of CRSFs in the wind-fed accreting X-ray pulsar systems like Vela X-1 have never been studied in detail. With the different companion star and accreting processes, the relationships between the CRSFs and the continuum spectral properties may be different or have some common characteristics. In this work, I will use all the available observed data on Vela X-1 collected by INTEGRAL from 2003 -- 2011 to systematically study the variations of spectral properties of this source in hard X-ray bands. The main science aims will try to search for the CRSFs in the hard X-ray spectra of 3 -- 200 keV, and study its variations with long-term monitoring data. So that I can detailedly study the relations between CRSFs and continuum spectral properties in Vela X-1 in different luminosity ranges. These correlations will help us to understand the accretion geometry and processes near the surface of the strongly magnetized neutron star in a wind-fed accreting X-ray binary.

The paper is organized as follows. In \S 2, the INTEGRAL observations is briefly introduced. The spectral properties of Vela X-1 in different accretion states are studied and cyclotron scattering lines are searched for in different luminosity ranges in \S 3, and possible correlations between CRSFs and spectral parameters of Vela X-1 in wide luminosity ranges are approached. In \S 4, the spectral variations of Vela X-1 over its orbital phases are also presented. In the last section \S 5, a brief summary and discussion are presented.

\section{INTEGRAL Observations}

\begin{figure*}
\centering
\includegraphics[angle=0,width=14cm]{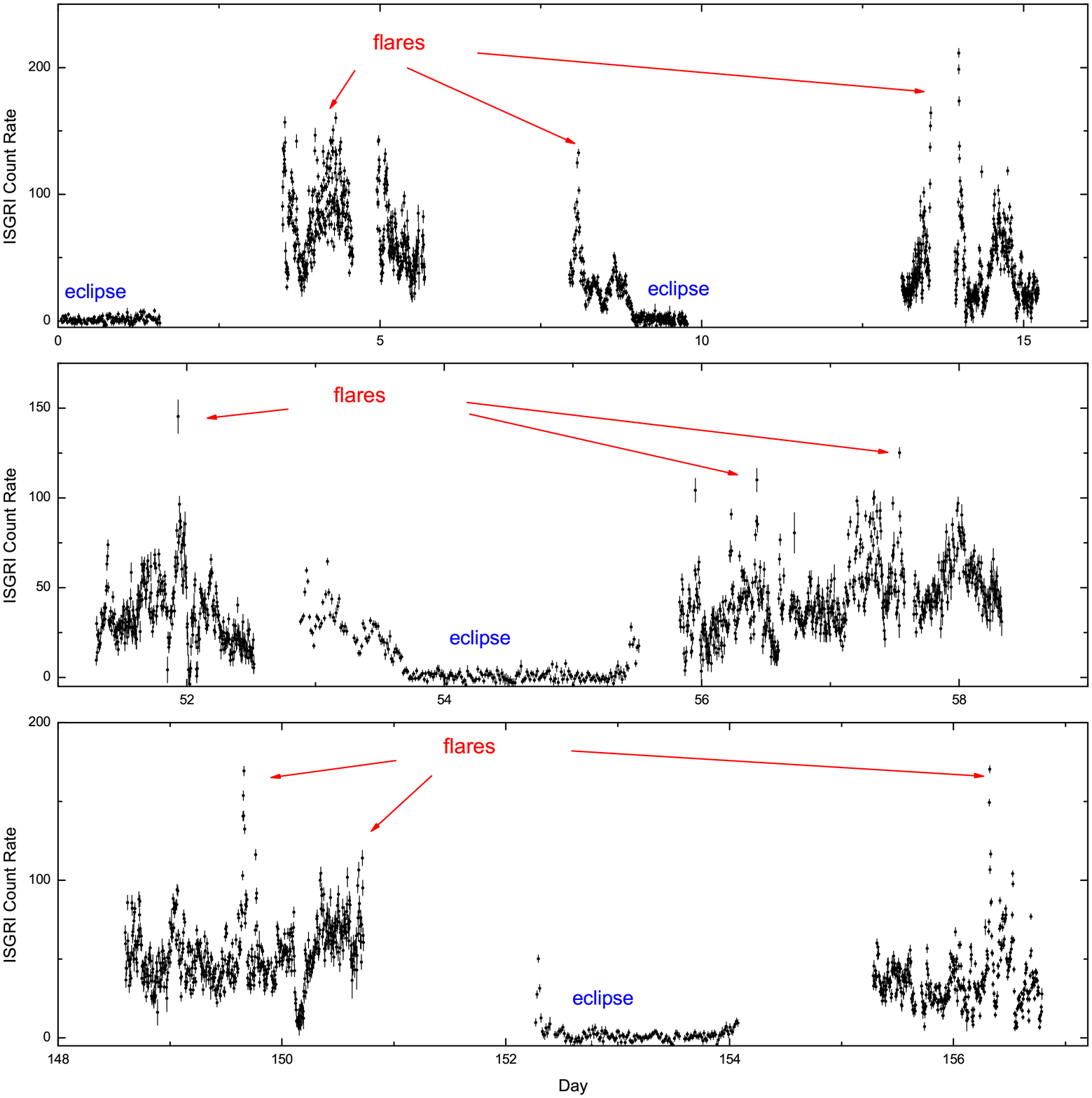}
\caption{The hard X-ray count rate variations of Vela X-1 from 20 -- 40 keV derived by long-term monitoring observations by IBIS-ISGRI from 2009 November -- 2010 April. Vela X-1 is a highly variable X-ray source with strong eclipses. In the observed interval, I detected four eclipses with a duration of $\sim 1.7$ days. Vela X-1 generally has a mean count rate of $\sim 40 - 50$ cts/s. Sometimes, it shows the flare states with a mean rate higher than 100 cts/s.}
\end{figure*}

Vela X-1 was observed with frequently pointing observational surveys on Vela region from 2003 -- 2011 by the INTEGRAL satellite. I mainly use the observational data obtained by the INTEGRAL Soft Gamma-Ray Imager (IBIS-ISGRI, Lebrun et al. 2003) which has a 12' (FWHM)
angular resolution and arcmin source location accuracy in the energy band of 15 -- 200 keV. JEM-X aboard INTEGRAL is the small X-ray telescope (Lund et al. 2003) which can be used to constrain the continuum spectrum below 20 keV combined with IBIS.

In this work, I use the available archival data for the IBIS observations
where Vela X-1 is within $\sim 10$ degrees of the pointing
direction. The INTEGRAL observations showed a long-term monitoring on Vela X-1 from 2003 -- 2011. The total observations have 1670 science windows (duration of each science window is about 2000 s). The archival data used in our work are available from the INTEGRAL Science Data Center (ISDC). The analysis is done with the standard INTEGRAL off-line
scientific analysis (OSA, Goldwurm et al. 2003) software, ver. 10.

Individual pointing observations processed with OSA 10 are mosaicked to create sky images for the source detection. I have used the energy range of 20 -- 40 keV by IBIS for source detection. Fig. 1 also shows a light curve sample of Vela X-1 obtained by IBIS. Vela X-1 is a strongly eclipsing binaries. The eclipses are observed by IBIS with the count rate decreasing to near zero, which lasts about 1.7 days consistent with the  report by Watson \& Griffiths (1977). Other than eclipses, Vela X-1 is also a highly variable hard X-ray source. From the hard X-ray count rate variations of Vela X-1 presented in Fig. 1, I define three accreting states based on the average luminosities: flare states with the mean count rate higher than 100 cts/s; active states, the mean count rate in the range of $\sim 30 - 100$ cts/s; low states with the mean count rate below $\sim 30$ cts/s (not including the eclipse data).

\section{Hard X-ray spectral properties of Vela X-1 in different accreting states}

\begin{figure*}
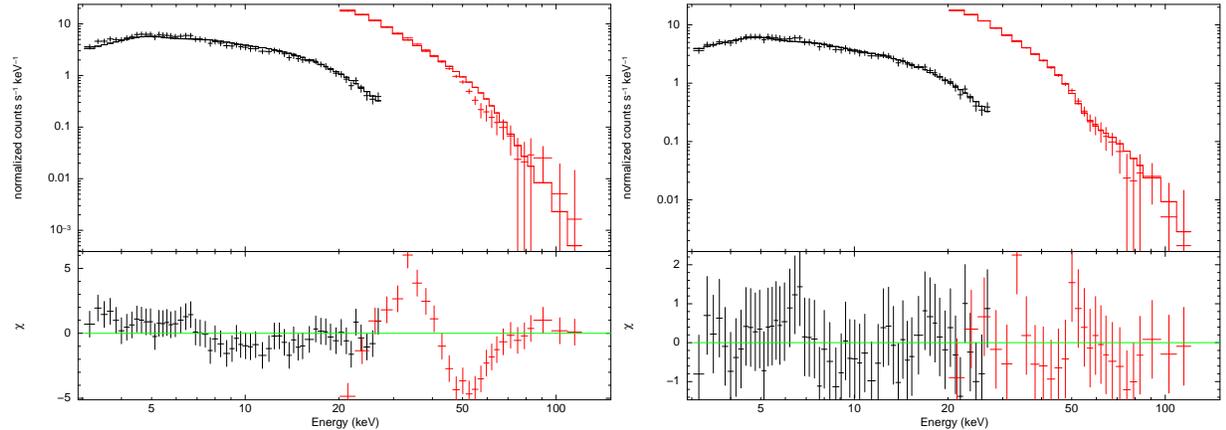

\includegraphics[angle=-90,width=8cm]{velax1_spec3820027_1.eps}
\includegraphics[angle=-90,width=8cm]{velax1_spec3820027.eps}
\caption{The hard X-ray spectrum sample from 3 --  200 keV of Vela X-1 in the flare state obtained by JEM-X and IBIS. {\bf Left} The spectrum is fitted with the cutoff power-law model.  {\bf Right} The spectrum is fitted with a cutoff power-law model plus a cyclotron scattering line around 55 keV.  }
\end{figure*}

\begin{figure*}
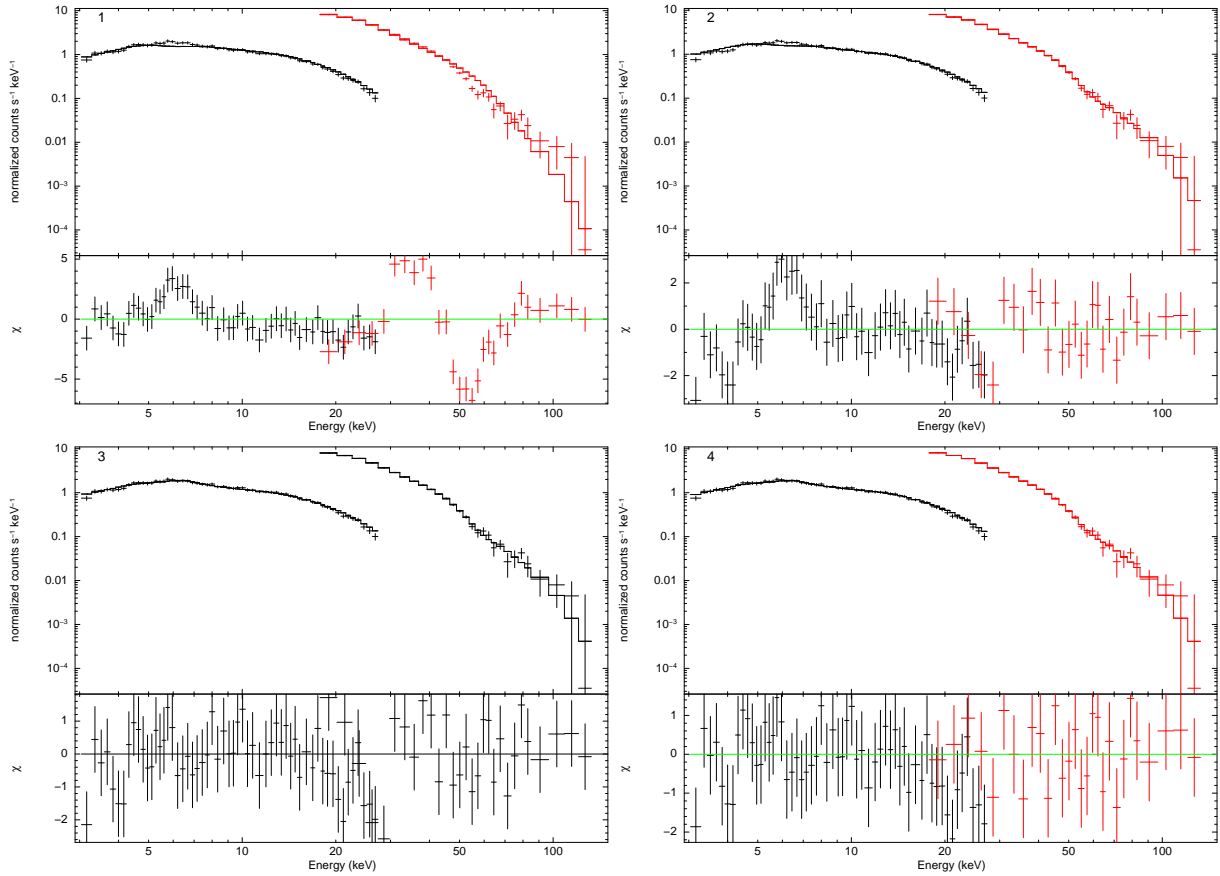

\includegraphics[angle=-90,width=8cm]{velax1_spe380_70_1.eps}
\includegraphics[angle=-90,width=8cm]{velax1_spe380_70_2.eps}
\includegraphics[angle=-90,width=8cm]{velax1_spe380_70_3.eps}
\includegraphics[angle=-90,width=8cm]{velax1_spe380_70_4.eps}
\caption{The hard X-ray spectrum sample from 3 --  200 keV of Vela X-1 in the active state obtained by JEM-X and IBIS. {\bf 1.} The spectrum is fitted with the cutoff power-law model. {\bf 2.} The spectrum is fitted with a cutoff power-law model plus a cyclotron scattering line around 56 keV. {\bf 3.} The spectrum is fitted with a cutoff power-law model plus a cyclotron scattering line at 56 keV with an Fe K$\alpha$ line at 6.4 keV. {\bf 4.} The spectrum is fitted with a cutoff power-law model plus two cyclotron scattering lines at $\sim 25$ and 56 keV with an Fe K$\alpha$ line at 6.4 keV.  }
\end{figure*}

\begin{figure*}
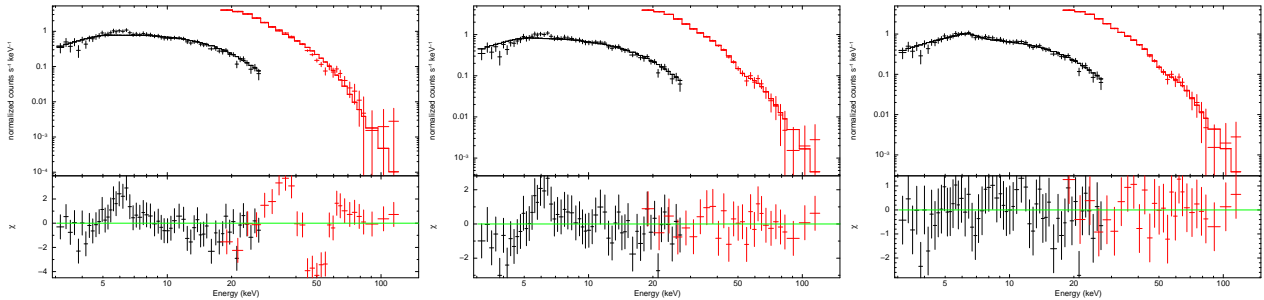

\includegraphics[angle=-90,width=5.5cm]{velax1_spe80_30_1.eps}
\includegraphics[angle=-90,width=5.5cm]{velax1_spe80_30_2.eps}
\includegraphics[angle=-90,width=5.5cm]{velax1_spe80_30_3.eps}
\caption{The hard X-ray spectrum samples from 3 --  200 keV of Vela X-1 in the low state the obtained by JEM-X and IBIS. {\bf Left} The spectrum are fitted with the cutoff power-law model. {\bf Middle} The spectrum is fitted with a cutoff power-law model plus a cyclotron scattering line around 50 keV. {\bf Right} The spectrum is fitted with a cutoff power-law model plus a cyclotron scattering line around 50 keV with an Fe K$\alpha$ line at 6.4 keV. }
\end{figure*}

The hard X-ray spectra of Vela X-1 from 3 -- 200 keV combined with JEM-X and IBIS are extracted for different accreting states. The cross-calibration studies on the JEM-X and IBIS/ISGRI detectors have been done using the Crab observation data, and the calibration between JEM-X and IBIS/ISGRI can be good enough within $\sim 6\%$ (see samples in Jourdain et al. 2008 and Wang 2013). In the spectral fittings, the constant factor between JEM-X and IBIS is set to be 1. The spectral analysis software package used is XSPEC 12.6.0q (Arnaud 1996).

Generally the hard X-ray spectrum of accreting X-ray pulsars like Vela X-1 can be described by a power-law model plus a high energy exponential rolloff (hereafter {\em cutoffpl}): $A(E)=KE^{-\Gamma}\exp(-E/E_{\rm cutoff})$. Other simple spectral models like the single power-law model and the thermal bremsstrahlung model are also applied to fit the spectra of Vela X-1. However these models cannot fit the hard X-ray spectra of Vela X-1 from 3 -- 200 keV, so I will use the cutoff power-law model (similar to Orlandini et al. 1998 and Kreykenbohm et al. 2008) in the following spectral analysis. After the continuum fittings on the spectrum, there still exist the dips in the residuals, like the absorption feature around 55 keV, which is the known cyclotron scattering line. So that I add the cyclotron scattering components to improve the spectral fittings. I also use the XSPEC model {\em cyclabs} to fit the cyclotron scattering line component (Mihara et al. 1990): $M(E)=\exp[-D_f{(W_fE/E_{cyc})^2 \over (E-E_{cyc})^2+W_f^2}]$. For some cases, the iron line at 6.4 keV is needed to fit the feature around 6 -- 7 keV in JEM-X spectra. I take a gaussian iron line profile in the fits: $A(E)=(K/\sqrt{2}\sigma_{Fe})\exp({-(E-E_{Fe})^2/2\sigma_{Fe}^2})$, where $E_{Fe}=6.4$ keV is the energy of iron line, $\sigma_{Fe}$ is the line width, $K$ is the line flux. In addition, a systematic uncertainty of $1\%$ for the IBIS-ISGRI energy channels is added in the spectral analysis.

\begin{table*}
%\tabletypesize{\scriptsize}
\scriptsize
\caption{The best fitted spectral parameters of Vela X-1 in three different luminosity ranges in the hard X-ray bands from 3 -- 200 keV. The spectra were fitted with different models as defined here: (1) the cutoff power-law model {\em cutoffpl}; (2) the cutoff power-law plus a cyclotron scattering line {\em cutoffpl*cyclabs}; (3) the cutoff power-law plus a cyclotron scattering line with an Fe K$\alpha$ line fixed at 6.4 keV; (4) the cutoff power-law plus two cyclotron scattering lines with an Fe K$\alpha$ line fixed at 6.4 keV. The width of the Fe K$\alpha$ line is also fixed to be zero in the fittings. The continuum flux is given in units of $10^{-9}$ erg cm$^{-2}$ s$^{-1}$, and the Fe line flux in units of $10^{-2}$ photons cm$^{-2}$ s$^{-1}$.}
% \setlength{\tabcolsep}{1.0mm}
%\tablewidth{0pt}
\begin{center}
\begin{tabular}{l c c c c c c c c c c l}
\hline \hline
model & $\Gamma$ & $E_{\rm cutoff}$ & $E_1$ &  $Width_1$ & $Depth_1$ & $E_2$ &$Width_2$& $Depth_2$  &Line flux & Flux & reduced $\chi^{2}\ (d.o.f)$ \\
  & & keV& keV& keV& & keV& keV& & & & \\
\hline
 &Flare  & & & & & & & & & \\
1& $-0.12\pm 0.03$ & 10.1$\pm 0.1$ & & & & & & & & $25.1\pm 0.4$& 4.0(82)\\
2& 0.14$\pm 0.04$ & $11.3\pm 0.4$& & & & 55.4$\pm 0.9$& 9.5$\pm 1.8$& 1.4$\pm 0.1$& & $24.2\pm 0.4$ & 0.6(79)\\
\hline
 &Active & & & & & & & & & \\
1& -0.34$\pm 0.02$ & 9.2$\pm 0.1$& & & & & & & & $9.2\pm 0.2$& 4.9(82)\\
2& -0.11$\pm 0.03$&11.1$\pm 0.2$& & & & 57.0$\pm 0.8$& 9.3$\pm 1.8$& 1.2$\pm 0.1$&  & $9.0\pm 0.2$& 1.7(79)\\
3& -0.17$\pm 0.03$ &10.9$\pm 0.2$& & & &56.9$\pm 0.8$& 9.2$\pm 1.8$& 1.2$\pm 0.1$&1.2$\pm0.2$ & $9.0\pm 0.2$& 1.1(78)\\
4&-0.14$\pm 0.03$&$11.1\pm 0.2$&25.5$\pm 0.9$& 3.7$\pm 1.7$ &0.06$\pm 0.01$&56.2$\pm 0.7$&9.3$\pm 1.8$&1.2$\pm0.1$ &1.2$\pm0.2$ & 8.9$\pm 0.2$ & 0.8(75)\\
\hline
 & Low & & & & & & & & & \\
1& -0.51$\pm 0.04$&8.4$\pm 0.1$& & & & & & & & $4.3\pm 0.3$& 2.8(82)\\
2&-0.24$\pm 0.05$&10.2$\pm 0.3$& & & &50.8$\pm 0.7$&7.9$\pm 1.7$&0.9$\pm 0.1$&  &$4.2\pm 0.3$& 1.1(79)\\
3&-0.31$\pm 0.06$&9.9$\pm 0.3$& & & &$50.7\pm 0.8$&$8.1\pm 1.7$&0.9$\pm 0.1$& 0.8$\pm0.2$ &$4.2\pm 0.4$& 0.7(78)\\
\hline
\end{tabular}
\end{center}
\end{table*}

%\begin{figure}
%\includegraphics[angle=0,width=9cm]{rev382_spec_t.eps}
%\caption{The spectral variations of Vela X-1 around a giant flare occurring on November 30, 2005. The start time of the light curve is MJD %53703.23.}
%\end{figure}
\begin{figure*}
\centering
\includegraphics[angle=0,width=16cm]{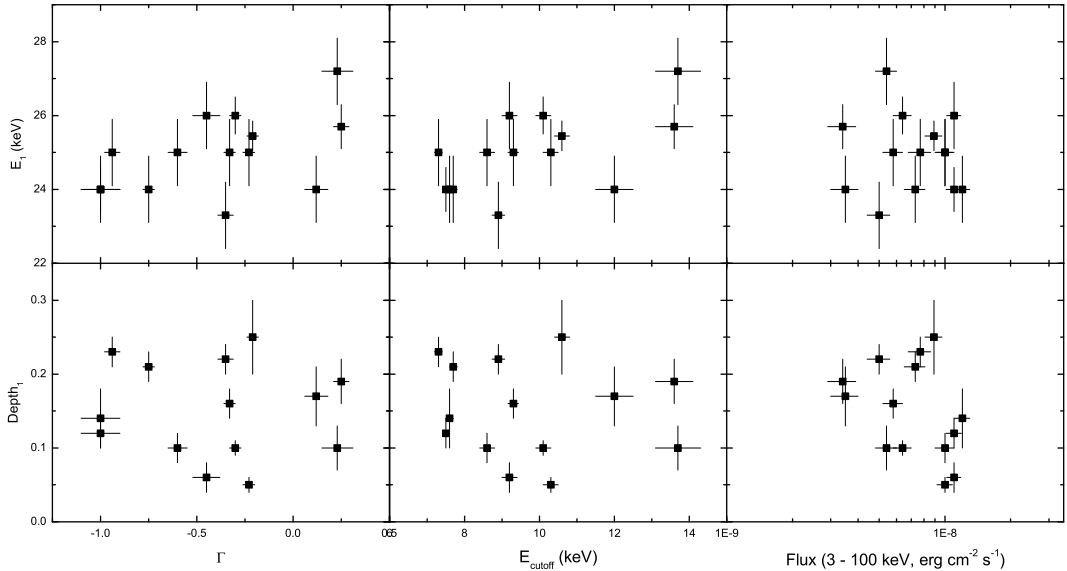}
\caption{The derived cyclotron energy $E_1$ and depth of the fundamental line versus three spectral parameters: hard X-ray flux in the range of 3 -- 100 keV, photon index $\Gamma$ and exponential cutoff energy $E_{\rm cutoff}$.  }
\end{figure*}

\begin{figure*}
\centering
\includegraphics[angle=0,width=16cm]{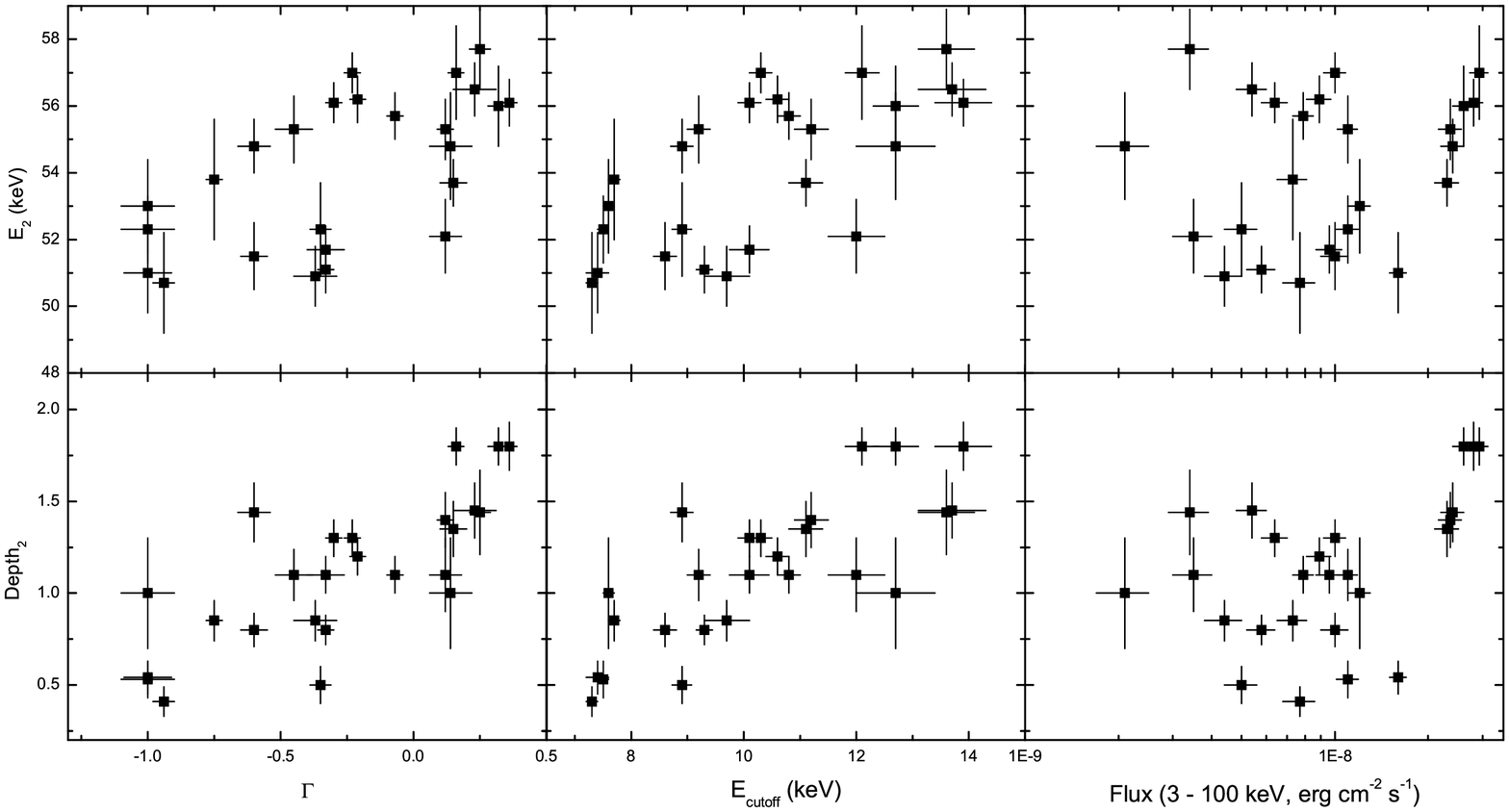}
\caption{The derived cyclotron energy $E_2$ and depth of the first harmonic versus three spectral parameters: hard X-ray flux in the range of 3 -- 100 keV, photon index $\Gamma$ and exponential cutoff energy $E_{\rm cutoff}$.}
\end{figure*}

\begin{figure*}
\centering
\includegraphics[angle=0,width=16cm]{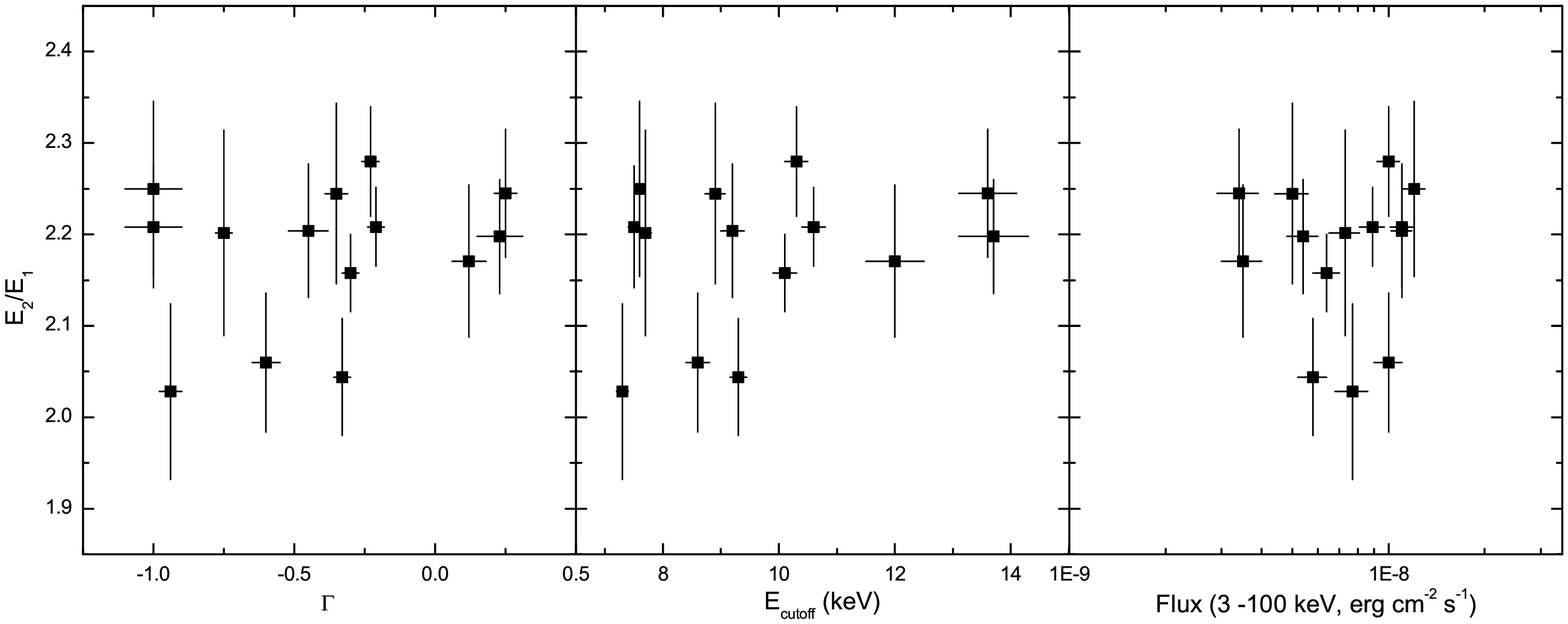}
\caption{The energy ratio of the first harmonic to the fundamental line $E_2/E_1$ versus three spectral parameters: hard X-ray flux in the range of 3 -- 100 keV, photon index $\Gamma$ and exponential cutoff energy $E_{\rm cutoff}$. The ratio has no correlations with hard X-ray flux, $\Gamma$ and the cutoff energy $E_{\rm cutoff}$.}
\end{figure*}

\begin{figure*}
\centering
\includegraphics[angle=0,width=15cm]{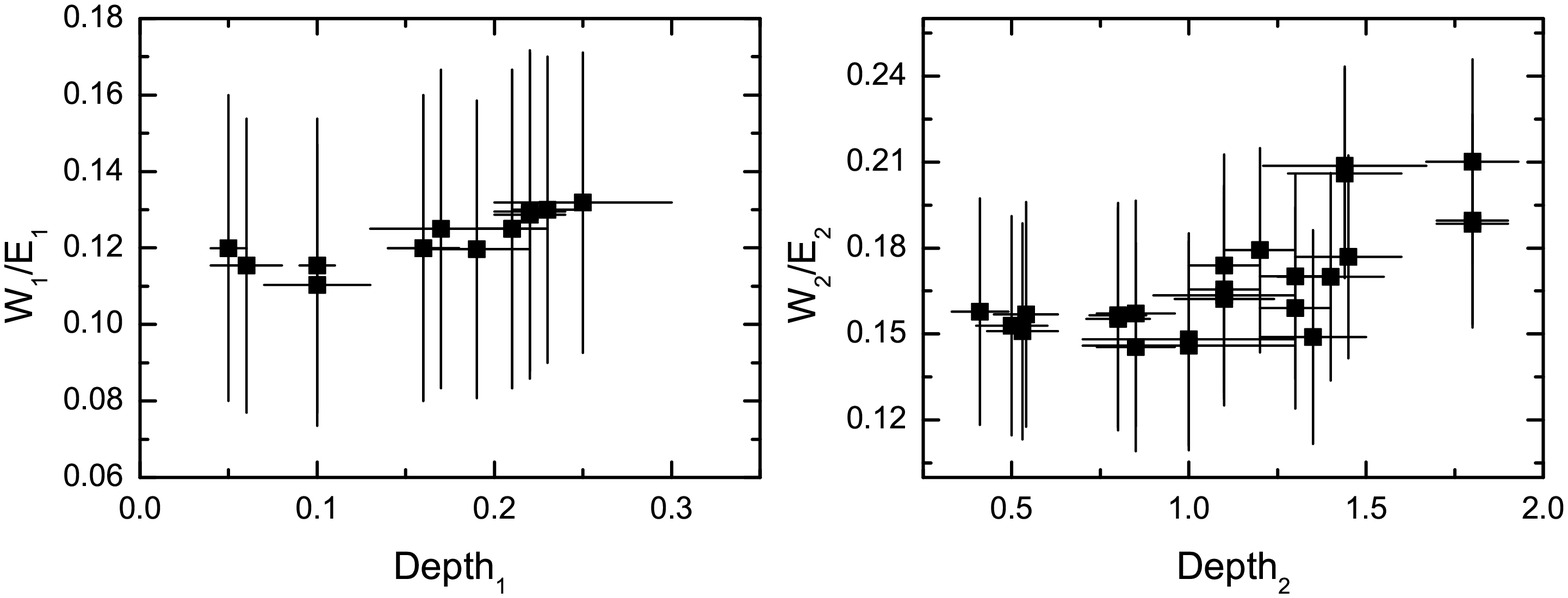}
\caption{The ratio of derived width ($W_i$) to energy $E_i$ of two cyclotron scattering lines of Vela X-1 versus their depth ($Depth_i$) respectively. A positive correlation between $W_2/E_2$ and $Depth_2$ is found, and there is no significant correlation of $W_1/E_1$ versus $Depth_1$.  }
\end{figure*}

I tried to obtain the spectra from 3 -- 200 keV of Vela X-1 in different luminosities for statistical studies like in \S 3.1. So that I combined different science windows in the similar IBIS count rate level into one spectral analysis. I then divided the IBIS count rates for all science windows (like IBIS count rates in Fig. 1) into several count rate levels according to the following definition: in the range of $> 0$ -- 100 cts/s (low and active states), the science windows are combined with a bin of 10 cts/s, i.e., 10 -- 20 cts/s; 70 -- 80 cts/s, and so on; for the count rate higher than 100 cts/s (flare states), the science windows are combined according to six count rate bins: 100 -- 120 cts/s, 120 -- 140 cts/s, 140 -- 160 cts/s, 160 -- 180 cts/s, 180 -- 200 cts/s and $>200$ cts/s.

Then the spectral extractions are carried out for each group of science windows in the data analysis. In Figs. 2 -- 4, I presented the samples of the hard X-ray spectrum for Vela X-1 in three accreting states. In Table 1, the best fitted spectral parameters of three accreting states are also displayed. In the followings, the hard X-ray spectral properties of Vela X-1 in three accreting states are described according to the analysis in this work.

In Fig. 2, one hard X-ray spectrum from 3 -- 200 keV in flare states is presented. The spectrum is first fitted with a power-law plus a high energy cutoff (the left panel of Fig. 2). There is an obvious absorption dip in the spectrum from 50 -- 60 keV, which should be due to the cyclotron resonance scattering. Thus in the right panel of Fig. 2, an cyclotron scattering line component is added into the spectral fittings. The fitted spectral parameters are collected and presented in Table 1. In flares states, the first harmonic of CRSFs in Vela X-1 varies from 51 -- 57 keV. But the fundamental line feature around 25 keV is not detected in flare states. The disappearance of the fundamental line is interesting and needs more studies. This non-detection can not be due to the low significance since in flare states, Vela X-1 was detected generally with a very high detection significance level $>400\sigma$ from 20 -- 40 keV. I also try to derive an upper limit for the depth for the fundamental line in the flare states, which is $\sim 0.05$ (2$\sigma$).

In the active states, two cyclotron scattering line components are detected in hard X-ray spectra of Vela X-1. Fig. 3 displays a hard X-ray spectrum of Vela X-1 in active states and the best fitted spectral parameters are also shown in Table 1. The spectral fitting processes are divided into four steps (also see Fig. 3 and Table 1): (1) the continuum fitted by a cutoff power-law model; (2) a resonant scattering line around 56 keV is added; (3) the Fe K$\alpha$ line at 6.4 keV is fitted; (4) there still exists an absorption line feature around 25 keV which is the fundamental line, then finally the spectrum is fitted with a cutoff power-law plus two resonant scattering lines with the fixed Fe K$\alpha$ line.

The reported fundamental cyclotron lines in active states are determined at the energies distributed from 22 -- 27 keV. The surface magnetic field of neutron star in Vela X-1 can be estimated by using the formula \beq [B/10^{12}{\rm G}]=[E_{\rm cyc}/11.6{\rm keV}](1+z), \enq where $E_{\rm cyc}$
is the energy of the fundamental line, here I take $E_{\rm cyc}=27$
keV, and $z$ is the gravitational redshift near the surface of the
neutron star. Then I obtain a magnetic field of $\sim 2.3(1+z)\times 10^{12}$ G for
the neutron star in Vela X-1. For the neutron star in Vela X-1, I can take the mass of 1.9 \ms (Quaintrell et al. 2003) with a
radius of 10 km, then I can estimate $B\sim 3\times 10^{12}$ G.

The first harmonic of CRSFs in Vela X-1 is determined at energies distributed from 50 -- 58 keV in active states. In addition, when compare fitted parameters of two cyclotron line features, I find that the fundamental line is generally narrower and shallower than those of the first harmonic in Vela X-1, i.e., the width of the fundamental line is around 4 keV, the depth is around 0.1 -- 0.3, and the width of the first harmonic is around 9 keV with the depth of $\sim 1$. This special feature may imply the large viewing angle with regard to the magnetic field in the case of Vela X-1 (see theoretical calculations by Nishimura 2011). The line width ratio $width_2/width_1\sim 2$ is still consistent with the predictions of theoretical Doppler broadening (M\'esz\'aros \& Nagel 1985).

In Fig. 4, I presented a hard X-ray spectrum in the low state of Vela X-1. The fitted spectral models and spectral parameters are also displayed in Table 1. In the low states, the fundamental cyclotron scattering line of CRSFs is not detected in the average spectra of Vela X-1, which may be due to the low detection significance levels on the source (lower X-ray flux levels). The determined first harmonic line energies are around 50 -- 57 keV, line width and depth also around 8 keV and 1 -- 2 respectively, which are still similar to the fitted parameters in the active states.

\subsection{Variations of CRSFs versus continuum spectral parameters}

The production of cyclotron resonant absorption line features near the surface of the neutron star is complicated in physics, which will sensitively depend on accretion states and accretion geometry (e.g., Araya \& Harding 1999; Nishimura 2011). In addition, the fundamental line energy may also change in different accretion luminosity states and the relativistic cyclotron scattering line energy would have non-harmonic line spacing (Araya \& Harding 2000). Therefore, detection of cyclotron scattering line features is not only used to identify a magnetized neutron star in binary systems but also important for studies on accretion process and geometry near the neutron star surface (Becker et al. 2012; Nishimura 2011). Variation pattern of the CRSFs and its physical origins need detailed studies with frequent observations. The correlation studies for the cyclotron resonant absorption line features should be helpful to understand the spectral properties and production mechanism of CRSFs near the surface of the magnetized neutron star.

In this subsection, using the INTEGRAL spectral analysis results in different accreting luminosities of Vela X-1, I will systematically study the relationships between cyclotron scattering lines and three continuum spectral parameters: hard X-ray flux, photon index and cutoff energy. Moreover, I not only show the fundamental line but also the first harmonic which is a stronger absorption feature in Vela X-1. In the followings, we used the Pearson product-moment correlation coefficient (hereafter $r$) to describe the correlation between spectral parameters, where $-1<r<1$, and larger $r$ values imply stronger correlation. Generally, there should exist the correlation between two parameters when $r>0.5$.

In Fig. 5, I present the relationship of the fundamental line energy $E_1$ and depth versus three continuum spectral parameters. $E_1$ shows no significant correlations with the X-ray flux, and the spectral parameters $\Gamma$ and $E_{\rm cutoff}$. But there may exist a very weak relation of $E_1$ versus the cutoff energy $E_{\rm cutoff}$ ($r=0.49$). This uncertainty may be due to a small number of data points so I will study this relation again in the following analysis. The depth has no relationship to three spectral parameters, may become lower when the flux is higher than $\sim 10^{-8}$ erg cm$^{-2}$ s$^{-1}$.

The first harmonic energy $E_2$ and depth versus the continuum spectral parameters are shown in Fig. 6. $E_2$ also has no relation with the X-ray flux, but shows the positive correlations with $\Gamma$ ($r=0.53$), and $E_{\rm cutoff}$ ($r=0.58$). Moreover, the absorption depth of the first harmonic also has the positive correlations with $\Gamma$ ($r=0.85$) and $E_{\rm cutoff}$ ($r=0.88$), but no relation to the accreting luminosity.

Furthermore, I present the relations between the energy ratio of the absorption lines $E_2/E_1$ and three continuum spectral parameters in Fig. 7. The ratio is independent of the X-ray flux, and $\Gamma$ and $E_{\rm cutoff}$ (maybe due to a small number of data points).  The derived energy ratio is generally higher than 2, suggesting the absorption line forming region with a large polar cap or of a greater height, so that the fundamental line would be formed at a higher site and the first harmonic primarily around the bottom (see simulations by Nishimura 2011).

Coburn et al. (2002) found that there exists a positive correlation
between absorption depth and ratio of width to energy in phase averaged spectra of the accreting X-ray pulsars. This indicates
that the accretion geometry of this accreting X-ray pulsar is a tall cylindrical shape accretion column rather than a flat
coin shape (Kreykenbohm et al. 2004). In Fig. 8, I also presented the
ratio of the width to energy of both $E_{1}$ and $E_{2}$ versus
the corresponding depth by different observations on Vela X-1. The weak positive correlation between $W_2/E_2$ and $Depth_2$ is found ($r=0.55$), I could not find the relationship between $W_1/E_1$ and $Depth_1$ (the ratio of $W_1/E_1$ is nearly constant with large uncertainties). So the properties of cyclotron resonance scattering features like $E_2$ suggested that Vela X-1 should have a cylindrical column accretion geometry. These correlations will be studied again for the further check in the following analysis.

%\section{Correlation studies of cyclotron resonance spectral features}
\section{Spectral variations over the orbital phases}

\begin{figure*}
\includegraphics[angle=0,width=15cm]{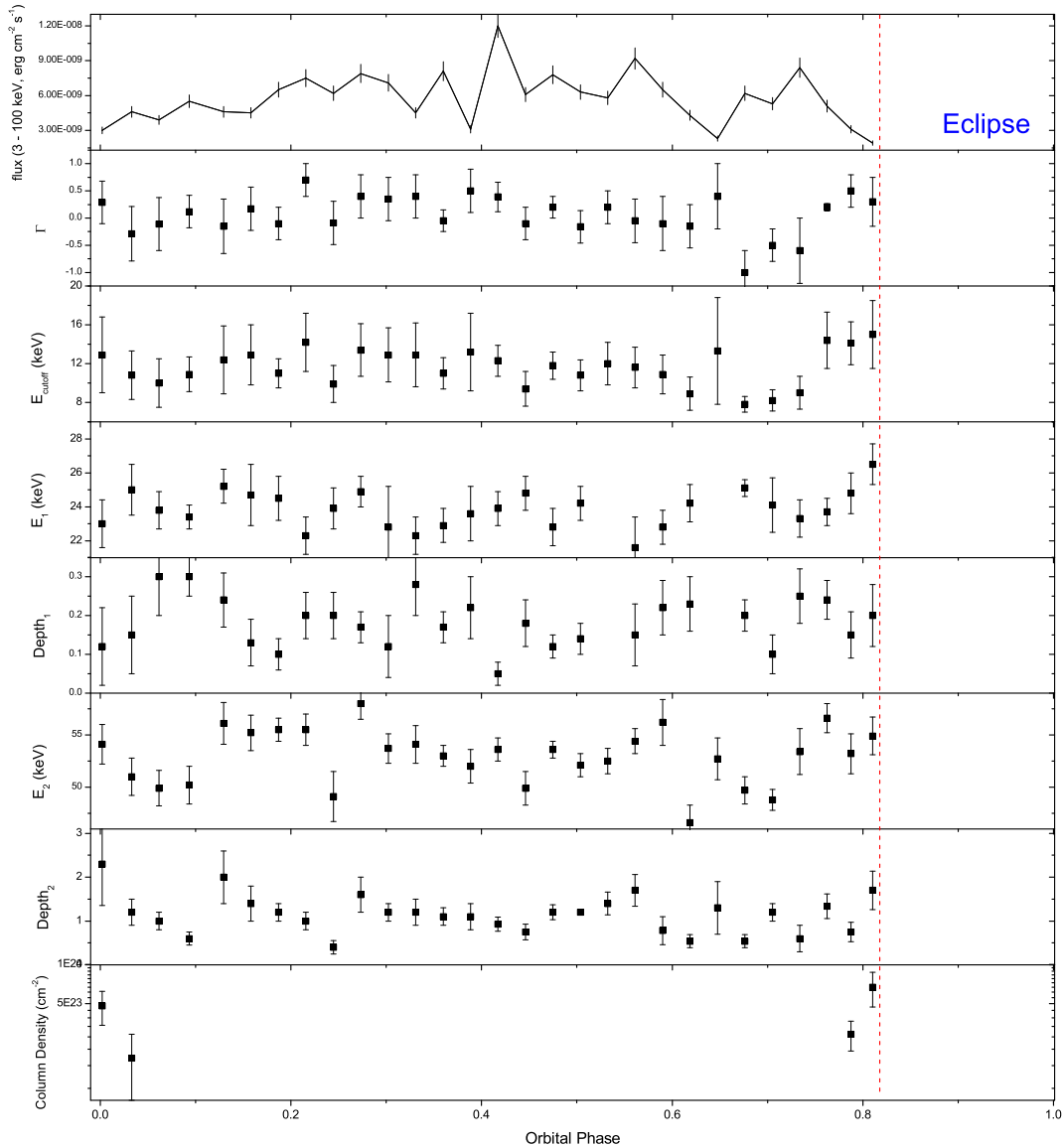}
\caption{The spectral variations of Vela X-1 over the whole orbital phase. No spectral information is derived during the eclipse phase of Vela X-1. The starting point is derived by the time just after the eclipse, which is taken as MJD 52975.075 here.}
\end{figure*}

With a long-term monitoring observation on Vela X-1 with INTEGRAL, I try to derive the spectral variations versus its orbital phase (an orbital period of $\sim 8.96$ days). There exist strong eclipses in the orbital phase of Vela X-1 (also see Fig. 1). During the eclipses (about 1.7 day), INTEGRAL did not detect the source, no spectral information can be obtained. For the other orbital phases, I divided the phase into 30 bins. And then the science windows of INTEGRAL observations are re-combined according to time intervals of different orbital phase bins. The data analysis was re-done for all science windows in each phase bin, and I extracted the spectrum for the data of each phase bin with JEM-X and IBIS.

Similar to \S 3, I fit the hard X-ray spectrum with the cutoff power-law model plus cyclotron scattering line features, to obtain the best fitted spectral parameters. For the orbital phase just before and after the eclipse, the cutoff power-law model cannot fit the spectrum below 5 keV very well. An additional hydrogen absorption component is required to fit the spectrum below 5 keV. The large column density values of $\sim (1-6)\times 10^{23}$ cm$^{-2}$ suggest the enhanced wind density just before and after the eclipses. In the other orbital phases, the absorption component is not needed in the spectral fittings, implying an upper limit of $\sim 10^{23}$ cm$^{-2}$ ($2\sigma$) for the column density. Kaper et al. (1994) suggested the absorption component structures resulting from the presence of a photo-ionization wake in Vela X-1 by optical spectroscopy of the companion. This absorption density variation versus orbit was also recently reported by MAXI (Doroshenko et al. 2013), which suggested that there exists a denser stream-like region
trailing the neutron star of Vela X-1. However the absorption density variation versus orbit (see Fig. 9) derived by INTEGRAL still support the standard smooth-wind model (Castor et al. 1975).

\begin{figure}
\centering
\includegraphics[angle=0,width=8cm]{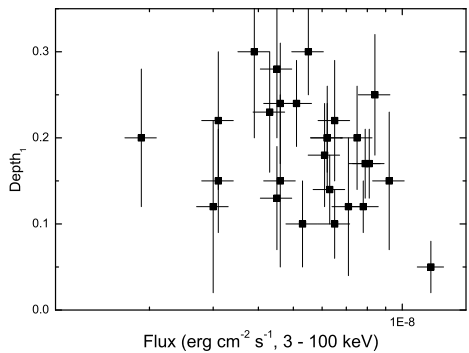}
\caption{The derived depth of the fundamental line versus hard X-ray flux in the range of 3 -- 100 keV.  }
\end{figure}

Finally, I derived all best fitted spectral parameters for the data in 30 orbital phase bins. In Fig. 9, the variations of the continuum spectral properties and two cyclotron scattering line features with the orbital phase are presented together. The parameters of both cyclotron scattering line features shows variations relative to continuum spectral parameters over the orbital phases. So similar to the correlation studies in \S 3.1, in this part, I also presented the relationships between cyclotron line features and three continuum spectral parameters.

For the fundamental line of Vela X-1, the energy $E_1$ and depth have no correlations to the continuum parameters, flux, $\Gamma$ and the cutoff energy $E_{\rm cutoff}$. In Fig. 10, I only presented the relation of depth versus the X-ray flux. The depth shows no correlation with hard X-ray flux ($r=0.29$). In the higher flux region like above $\sim 10^{-8}$ erg cm$^{-2}$ s$^{-1}$, the depth becomes significant smaller than the lower flux region, which may indicate no detection of the fundamental line in the flare states (\S 3).

The fitted parameters of the first harmonic of CRSFs versus X-ray flux, $\Gamma$ and $E_{\rm cutoff}$ are displayed in Fig. 11. Both the cyclotron energy $E_2$ and absorption depth have no relation to hard X-ray flux. While $E_2$ has strong positive correlations with photon index $\Gamma$ ($r=0.56$) and the cutoff energy $E_{\rm cutoff}$ ($r=0.76$). The absorption depth shows no significant correlation to $\Gamma$, and a very weak positive correlation with $E_{\rm cutoff}$ ($r=0.49$). The correlations $E_2$ versus $E_{\rm cutoff}$, and depth versus $E_{\rm cutoff}$, support the idea that the high energy cutoff in the hard X-ray spectra of accreting X-ray pulsars is attributed to the cyclotron resonant scattering.

Fig. 12 also presents the ratio $E_2/E_1$ versus three continuum spectral parameters. The ratio is still generally larger than 2, and shows no relation to the hard X-ray flux. But $E_2/E_1$ shows the positive correlation with photon index $\Gamma$ ($r=0.52$) and the cutoff energy $E_{\rm cutoff}$ ($r=0.56$).

In Fig. 13, the ratios of width to line energies versus the absorption depth for two cyclotron scattering line features are studied again. Though the large uncertainties still exist, I can still find the positive relationships between the ratio and depth for each cyclotron line: $W_1/E_1$ versus $Depth_1$ ($r=0.63$) and $W_2/E_2$ versus $Depth_2$ ($r=0.69$). With more data analysis, the assumption of a cylindrical column accretion geometry near the surface of the magnetized neutron star in Vela X-1 is confirmed.

\begin{figure*}
\centering
\includegraphics[angle=0,width=16cm]{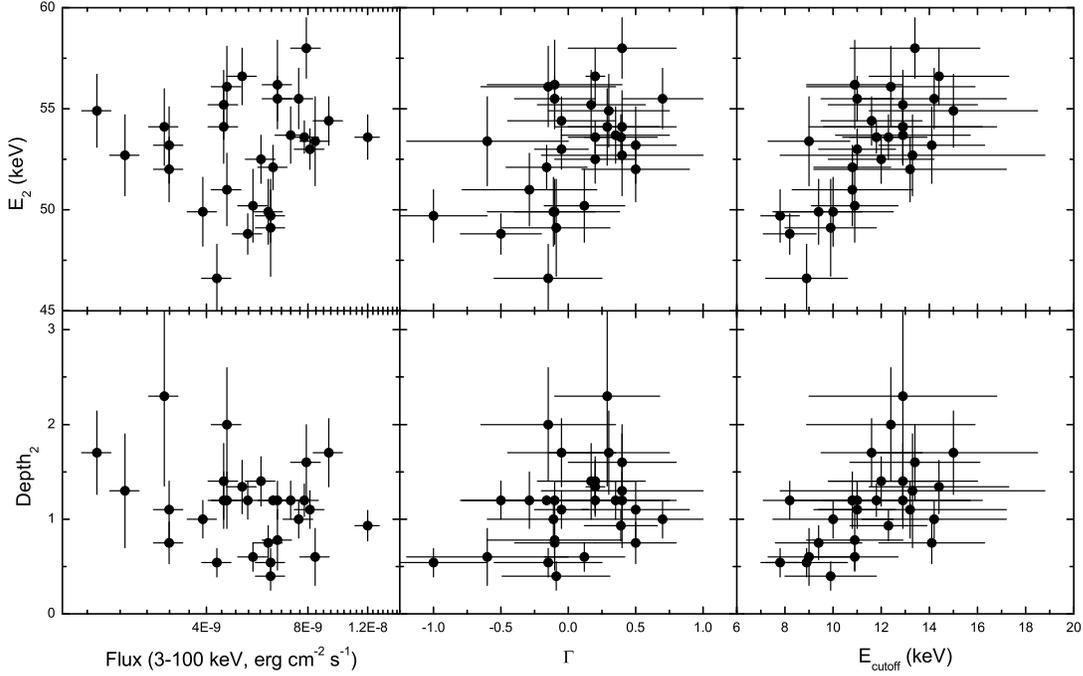}
\caption{The derived cyclotron energy $E_2$ and depth of the first harmonic versus three spectral parameters: hard X-ray flux in the range of 3 -- 100 keV, photon index $\Gamma$ and exponential cutoff energy $E_{\rm cutoff}$.}
\end{figure*}

\begin{figure*}
\centering
\includegraphics[angle=0,width=16cm]{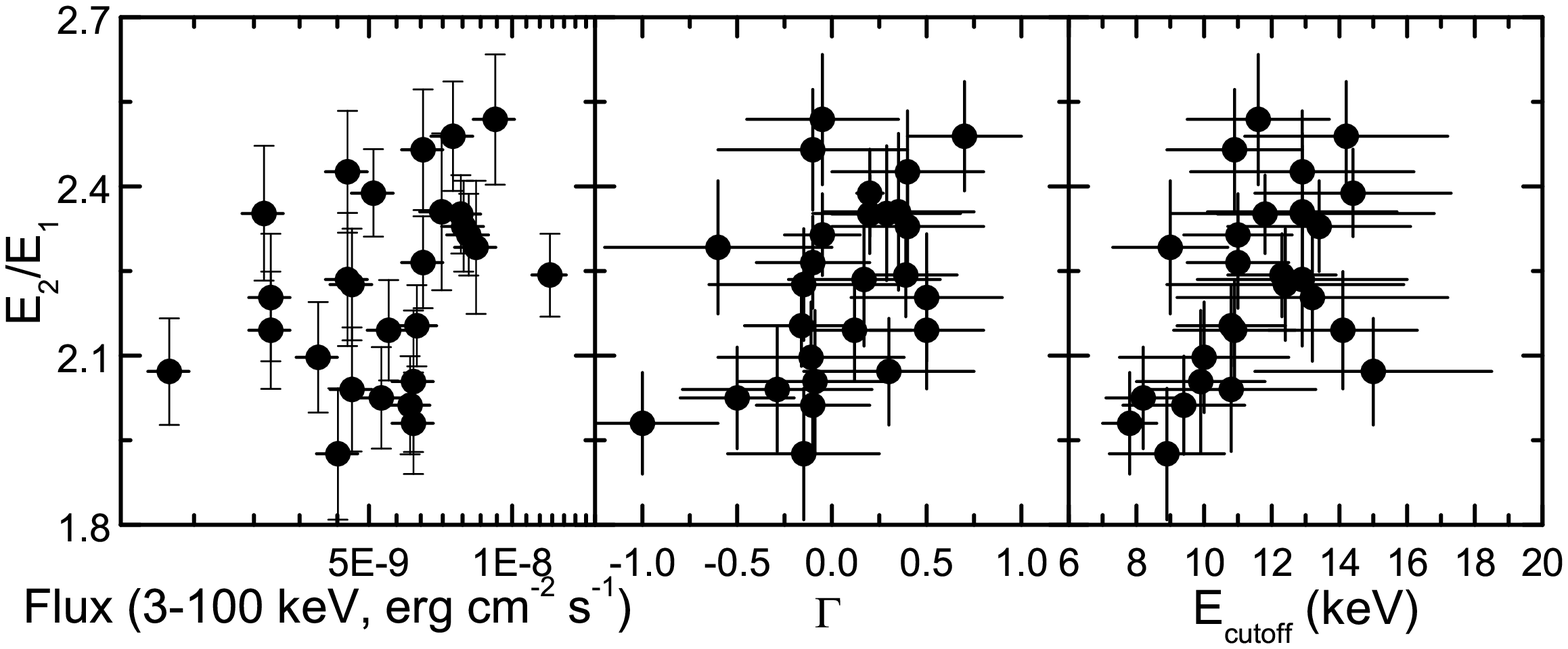}
\caption{The energy ratio of the first harmonic to the fundamental line $E_2/E_1$ versus three spectral parameters: hard X-ray flux in the range of 3 -- 100 keV, photon index $\Gamma$ and exponential cutoff energy $E_{\rm cutoff}$. The ratio has no correlation with hard X-ray flux, but shows the positive correlations with $\Gamma$ and the cutoff energy $E_{\rm cutoff}$.}
\end{figure*}

\begin{figure*}
\centering
\includegraphics[angle=0,width=15cm]{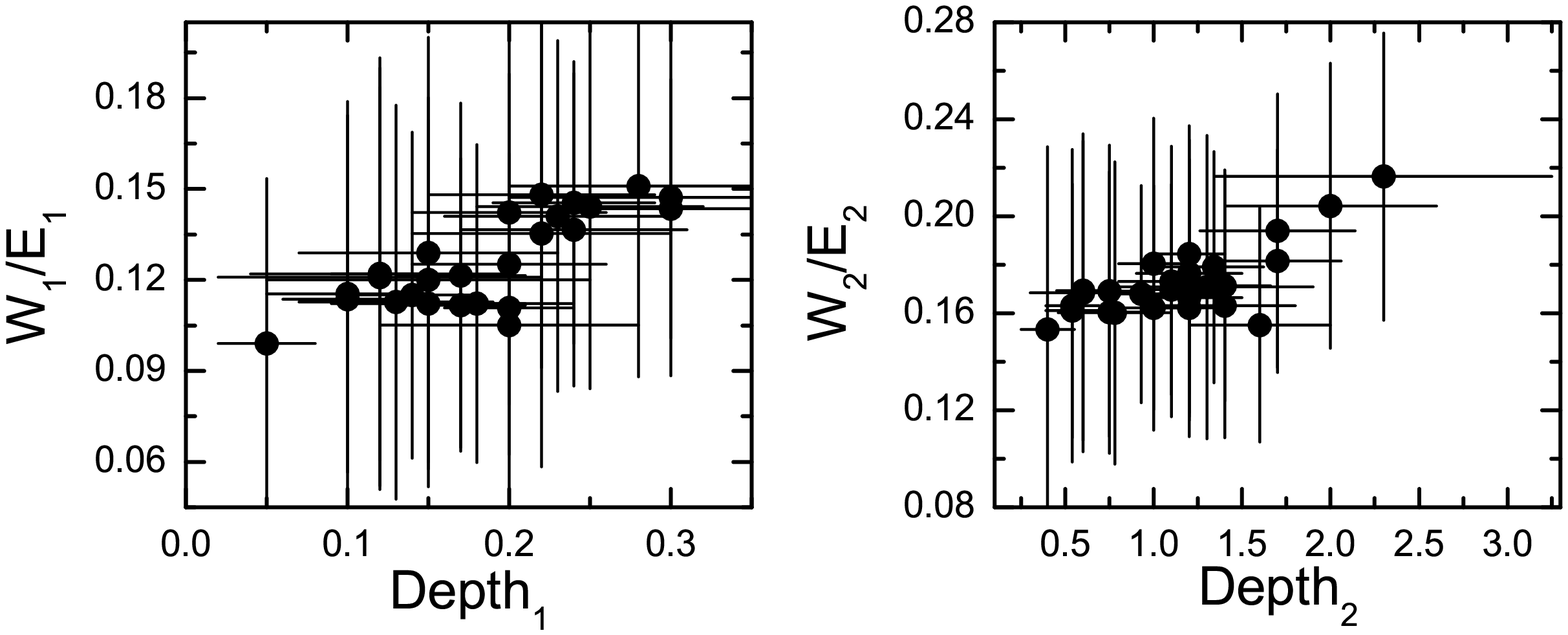}
\caption{The ratio of derived width ($W_i$) to energy $E_i$ of two cyclotron scattering lines of Vela X-1 versus their depth ($Depth_i$) respectively. A positive correlation between $W_2/E_2$ and $Depth_2$ is found, and there also exists a positive correlation of $W_1/E_1$ versus $Depth_1$.  }
\end{figure*}

\section{Summary and discussions}

In this work, I have carried out a long-term hard X-ray monitoring observation on Vela X-1 using INTEGRAL from 2003 -- 2011. With detailed analysis on the spectral properties of Vela X-1 in different luminosity ranges and orbital phases, I found some new results which are summarized here as follows: \\
(1) The cyclotron resonant scattering features are detected in Vela X-1 for different accreting states. In the average spectra of Vela X-1, the fundamental absorption line varies from 22 -- 27 keV, and the first harmonic is determined at 47 -- 58 keV. I obtain a surface magnetic field of $\sim 3\times 10^{12}$ G for Vela X-1 using the reported fundamental line energy. \\
(2) The fundamental absorption line cannot be detected in the flare states, with an upper limit of $\sim 0.05\ (2\sigma)$ for the depth. Above a critical X-ray flux $\sim 1.2\times 10^{-8}$ erg cm$^{-2}$ s$^{-1}$, corresponding to a luminosity from 3 -- 100 keV of $\sim 5\times 10^{36}$ erg s$^{-1}$, only the first harmonic around 50 -- 57 keV could be detected in average spectra of Vela X-1. In the active states when Vela X-1 has a lower luminosity, both the fundamental absorption line and the first harmonic are detected. When hard X-ray luminosity is below $\sim 5\times 10^{35}$ erg s$^{-1}$, the fundamental absorption line again cannot be detected, maybe due to the lower detection significance levels on the source. \\
(3) The cyclotron line energies and depths in different accretion luminosities and orbital phases are studied. The fundamental line energy shows no significant correlations with X-ray luminosity, photon index and high energy cutoff. While the first harmonic energy shows the positive correlations with photon index and cutoff energy. These positive correlations suggest that the first harmonic dominates the effects on the spectral profiles of Vela X-1 in hard X-ray bands. The line energy ratio $E_2/E_1>2$ shows the weak positive correlation with photon index and cutoff energy. The absorption depth of the fundamental line has no significant correlations to flux, $\Gamma$ and $E_{\rm cutoff}$, but in the high flux region above $10^{-8}$ erg cm$^{-2}$ s$^{-1}$, the depth becomes much smaller. The absorption depth of the first harmonic shows no relation to X-ray flux, $\Gamma$ and $E_{\rm cutoff}$.\\
(4) There exist the positive correlations between the ratio of line width to energy and the depth for both cyclotron scattering lines. This supports that the persistent wind-fed accretion source Vela X-1 also has a cylindrical column accretion geometry similar to the other supergiant X-ray pulsar GX 301-2 (Kreykenbohm et al. 2004) and some Be X-ray transient sources during their outbursts (see Li et al. 2012).\\
(5) Change of the absorption density with the orbital phases in Vela X-1 is also detected. The enhanced absorption is observed near the eclipse, and the density variation profile versus orbit phases is still consistent with a standard smooth-wind model.

With the long-term observations, I confirmed CRSFs at about 25 and 55 keV in the hard X-ray spectra of Vela X-1. More importantly I found large variations of the fitted parameters of CRSFs (especially the cyclotron energies). These variations may relate to accretion process and geometry which could be connected to X-ray continuum properties, like luminosity, photon index or spectral cutoff energies. For some X-ray pulsars, the possible relationship between the energy of the fundamental line and the bolometric luminosity is reported. A negative correlation between the energy of the fundamental line and the luminosity is found in a Be transient 4U 0115+63 during the outbursts (Li et al. 2012), while a recent work by M\"uller et al. (2013) showed that there is no correlation between the fundamental line energy and luminosity in this source. However, the positive correlation between the cyclotron line energy and luminosity is found in other X-ray pulsars like Her X-1 (Vasco et al. 2011) and GX 304-1 (Klochkov et al. 2012). Becker et al. (2012) suggested a critical X-ray luminosity in these accreting X-ray pulsars which can explain the bimodal dependence of the CRSF centroid energy on X-ray luminosity. Moreover, a positive correlation between the energy of the fundamental line and photon index was also discovered in 4U 0115+63 during the giant outburst in 2008 (Li et al. 2012). In addition, collecting observation data of different X-ray pulsars, a positive correlation between the energy of the fundamental CRSF and the cutoff energy in the canonical X-ray pulsar continuum is found (Makishima et al. 1999; Kreykenbohm et al. 2002; Coburn et al. 2002). The physical origins of these correlation are still not well understood, but are generally believed to provide strong constraints on production models of CRSFs and accretion geometry near the surface of the magnetized neutron star in HMXBs.

The existence of the correlations between cyclotron scattering energies and spectral photon index and cutoff energies suggests that the spectral properties of accreting X-ray pulsar Vela X-1 would be strongly affected by the cyclotron resonance similar to other X-ray pulsars. Previous work on statistical relation between cutoff energies and fundamental line energies for ten accreting X-ray pulsars has inferred that the spectral cutoff should be primarily caused by the cyclotron resonance (Makishima et al. 1999; Kreykenbohm et al. 2002; Coburn et al 2002). These ten X-ray pulsar systems included Be X-ray transients and wind-fed supergiant X-ray binaries. This correlation was derived using different types of X-ray pulsars and different sources. For the single source case of Vela X-1, I still confirmed the correlation. In addition, these relations are more complicated due to existence of two cyclotron scattering lines. The fundamental line energy has no relation to the cutoff energy, but the first harmonic energy show the strongly positive correlation with the cutoff energy. The possible explanation would be the first harmonic is broader and deeper than the fundamental line in the spectrum of Vela X-1 and it is always significantly detected but the fundamental line some times disappears, like in the flare state. So the first harmonic will dominate the cause of the spectral cutoff in Vela X-1.

Generally the fundamental cyclotron resonance is thought to have a much larger electron-photon interaction
cross section than the higher harmonic resonances which will lead to a broader fundamental line with a larger depth (like the case of a typical X-ray pulsar 4U 0115+63, see Li et al. 2012). But some X-ray pulsars with longer spin periods also show the broader and deeper first harmonic like Vela X-1 and 4U 2206+54 (Wang 2009). Some physical conditions may cause this special case in Vela X-1. Nishimura (2011) modeled cyclotron lines by considering superposition of series of cyclotron line spectra emerging from
different heights of a line-forming region, and simulated the profiles of fundamental and first harmonic lines considering different viewing angles. The simulation results found that in the larger viewing angles, the expected fundamental line profile will be shallow and narrow, and also suggested that in the case of Vela X-1, the fundamental line tends to be
formed around a higher site at the larger line-forming region, while the first harmonic line is formed primarily around the
bottom. Thus the predicted energy ratio of $E_2/E_1$ is higher than harmonic ratio 2.

In the analysis on the cyclotron line features of Vela X-1 in different luminosities, I have found that above a critical hard X-ray luminosity $\sim 5\times 10^{36}$ erg s$^{-1}$ from 3 -- 100 keV, the fundamental line cannot be detected for smaller absorption depth. This feature is also detected in the giant outbursts in a Be/X-ray pulsar A0535+26 when its X-ray luminosity higher than $\sim 10^{37}$ erg s$^{-1}$ (Maisack et al. 1997). Non-detection of the fundamental line in high luminosity ranges may suggest that accreting radiation in Vela X-1 would be strongly beamed (Poutanen \& Gierli\'nski 2003), and the pencil-like and fan-like beam patterns can explain the fundamental line disappearance in high luminosity ranges. These beamed patterns depend on the polar cap radius $R_p$ which is determined by the luminosity and the neutron star mass (see Becker 1998) \beq R_p\sim 0.4R_{NS}{M_\odot \over M_{NS}}{L_x\over 10^{37} {\rm erg s^{-1}}}, \enq where $M_{NS}$ is the mass of the neutron star, $R_{NS}$ is the radius of the neutron star. Then in high X-ray luminosity ranges of $>5\times 10^{36}$ erg s$^{-1}$ for Vela X-1 (the total accreting luminosity will be higher, $\sim 10^{37}$ erg s$^{-1}$, see Lutovinov \& Tsygankov 2009), one can derive $R_p>0.1 R_{NS}$, so that a fan-like beam pattern is expected in Vela X-1. According to the simulations by Nishimura (2011), in a fan-like beam pattern, the fundamental line tends to be shallower (consistent with the observed feature in Vela X-1, also see Fig. 10), while the first harmonic line tends to be deeper (also see Fig. 6).

It should be noted that the at present non-detection of the fundamental line in high X-ray luminosity ranges is only reported in two X-ray pulsar systems with relatively long spin period, Vela X-1 (283 s) and A0535+26 (103 s). However in some bright X-ray pulsars with short spin period, like 4U 0115+63 (3.6 s), even in the giant outburst peaks (X-ray luminosity $>5\times 10^{37}$ erg s$^{-1}$. see Li et al. 2012), the fundamental line around 10 -- 15 keV is detected, and shows very wide and deep feature. Thus the beam pattern transition could only occur in the long spin period X-ray pulsar systems. What differences exist between short spin period pulsars and long spin period ones in producing cyclotron scattering line features, and what mechanism (physics or geometry)? This issue would require further studies by observations and theories.

There are no correlations between X-ray luminosity and the cyclotron energies for both two lines in Vela X-1. This behavior is different from that discovered in Her X-1 and some Be X-ray transient sources during the outbursts though both Vela X-1 and Be X-ray transients (like 4U 0115+63) have the similar accretion geometry. In Be X-ray transients, X-ray luminosity plays a critical role in affecting the variations of the fundamental line energy (also see discussions in Becker et al. 2012). While in the case of Vela X-1, the variations of cyclotron scattering line energies of both fundamental and first harmonic would not be caused by the change of accretion luminosity but by some other physical properties. This difference in Vela X-1 and Be X-ray transients and Her X-1 (with a low-mass companion star) may be due to different type companion stars and accreting processes. Of course the variation range of observed luminosities in Vela X-1 is still smaller than those in Be X-ray transients, which may make it difficult that the present data are used to study the relation of cyclotron line energy versus very wide X-ray luminosity range in Vela X-1. Therefore, the variations and correlations of cyclotron resonance scattering features in Vela X-1 and other wind-fed supergiant X-ray pulsars need further studies both in observations and theoretical work.

\section*{Acknowledgments}
I am grateful to the referee for the fruitful and suggestive comments to greatly improve the manuscript. This paper is based on observations of
INTEGRAL, an ESA project with instrument and science data center
funded by ESA member states. W.W. is supported by the National Natural Science
Foundation of China (No. 11073030).

\end{document}